\documentstyle[12pt,aasms4]{article}

\slugcomment{}
\lefthead{}
\righthead{}

\begin{document}

\title
{Measurements of Anisotropy in the Cosmic Microwave Background Radiation\\
    at 0\fdg5 Scales\\
    near the Stars HR5127 and Phi Herculis}

\author{S.~T.~Tanaka\altaffilmark{1,2},
A.~C.~Clapp\altaffilmark{1,2,4},
M.~J.~Devlin\altaffilmark{1,2,6},
N.~Figueiredo\altaffilmark{1,3,7},
J.~O.~Gundersen\altaffilmark{1,3},
S.~Hanany\altaffilmark{1,2},
V.~V.~Hristov\altaffilmark{1,2},
A.~E.~Lange\altaffilmark{1,4},
M.~A.~Lim\altaffilmark{1,3},
P.~M.~Lubin\altaffilmark{1,3},
P.~R.~Meinhold\altaffilmark{1,3},
P.~L.~Richards\altaffilmark{1,2},
G. F. Smoot\altaffilmark{1,2,5},
and
J. Staren\altaffilmark{1,3}}

\altaffiltext{1}{NSF Center for Particle Astrophysics,
	Berkeley, CA 94720.}
\altaffiltext{2}{Physics Department,
	University of California at Berkeley,
	Berkeley, CA 94720}
\altaffiltext{3}{Physics Department,
	University of California at Santa Barbara,
	Santa Barbara, CA 93106}
\altaffiltext{4}{Division of Physics, Mathematics and Astronomy,
	California Institute of Technology,
	Pasadena, CA 91125}
\altaffiltext{5}{Physics Division, Lawrence Berkeley Laboratory,
	Berkeley, CA 94720}
\altaffiltext{6}{present address: Department of Physics,
	Princeton University,
	Princeton, NJ 08544}
\altaffiltext{7}{Escola Federal de Engenharia de Itajub\'{a},
	37500-000 Itajub\'{a}, MG, Brazil;
	Instituto Nacional de Pesquisas Espaciais, 12201-970
	S\~{a}o Jos\'{e} dos Campos, SP, Brazil}
\begin{abstract}
We present measurements of cosmic microwave background (CMB)
anisotropy near the stars HR5127 and Phi Herculis
from the fifth flight of the
Millimeter-wave Anisotropy eXperiment (MAX).
We scanned 8\arcdeg\ strips of the sky with
an approximately Gaussian 0\fdg5 FWHM beam
and a 1\fdg4 peak to peak sinusoidal chop.  The instrument has four
frequency bands centered at 3.5, 6, 9, and 14~cm$^{-1}$.
The IRAS 100~\micron\ map predicts that these two region
have low interstellar dust contrast.
The HR5127 data
are consistent with CMB anisotropy.
The Phi Herculis data, which were measured at lower flight altitudes,
show time variability at 9 and 14~cm$^{-1}$
which we believe to be due to atmospheric emission.
However, the Phi Herculis data at 3.5 and 6~cm$^{-1}$ are essentially
independent of this atmospheric contribution
and are consistent with CMB anisotropy.
Confusion from galactic foregrounds is unlikely based on the
spectrum and amplitude of the structure at these frequencies.
If the observed HR5127 structure and the atmosphere independent
Phi Herculis structure are attributed to CMB anisotropy, then
we find
$\Delta T/T = \langle\frac{l(l+1)C_l}{2\pi}\rangle^{1/2} =
1.2^{+0.4}_{-0.3}\times 10^{-5}$ for HR5127 and
$1.9^{+0.7}_{-0.4}\times 10^{-5}$
for Phi Herculis in the flat band approximation.
The upper and lower limits represent a 68\% confidence interval
added in quadrature with a 10\% calibration uncertainty.
\end{abstract}


\section{Introduction}

Measurements of the anisotropy of the cosmic microwave
background (CMB) provide an effective method for constraining
models of large scale structure formation.
Standard inflationary models predict a Doppler peak at angular
scales $\lesssim$1\arcdeg.
The Millimeter-wave Anisotropy eXperiment (MAX) measures
anisotropy on 0\fdg5 angular scales.
MAX has presented CMB results for six sky regions
from four previous flights
(\cite{fis92,als92,mei93a,gun93,cla94,dev94,fis95}).
Several other ground-based and balloon-borne experiments have also
reported detections at these angular scales
(\cite{deb94,gun95,net95,ruh95,che95}).
In this paper we report on two new CMB observations
from the fifth flight of MAX (MAX5).

\section{Instrument}

The instrument has been described in detail elsewhere
(\cite{fis92,als92,mei93b}).
MAX is an off-axis Gregorian telescope and a bolometric photometer
mounted on an attitude-controlled balloon platform.
The telescope has a 1~m off-axis parabolic primary with an elliptical
secondary which sinusoidally modulates the beam 1\fdg4 peak-to-peak
on the sky parallel to the horizon at
5.4~Hz.
The underfilled optics provide an approximately Gaussian
0\fdg5~FWHM beam.
For MAX5 an adiabatic demagnetization refrigerator cooled the
single-pixel, four-band photometer to 85~mK.
The four frequency bands were centered
at 3.5, 6, 9, and 14~cm$^{-1}$ with respective fractional bandwidths
0.5, 0.5, 0.4, and 0.2.  The ratio of
2.726~K thermodynamic temperature
to antenna temperature in each frequency band is
1.62, 2.50, 6.66, and 38.7, respectively.

\section{Observation}

The instrument was launched from the National Scientific Balloon
Facility in Palestine, Texas at 1.16~UT 1994 June~20.
The altitude during the observations drifted from 35~km to 32~km.
We observed CMB anisotropies in three sky regions.
Each CMB observation consists of constant velocity scans in azimuth
of $\pm$4\arcdeg\ while optically tracking a target star such that
sky rotation pivots the observed area about the star.
The angular offset between the center of the scan and the target star
is 0\fdg55 so that at the center of the scan the target star is in
one lobe of the chopped beam pattern.
The three MAX5 target stars are HR5127, Phi Herculis, and Mu Pegasi.
Lim et al. (1995) report on the Mu Pegasi observation.
We report here
the results for the sky regions near HR5127
($\alpha$~=~13$^{\rm h}$37\fm7, $\delta$~=~36\arcdeg19\arcmin)
and Phi Herculis
($\alpha$~=~16$^{\rm h}$08\fm6, $\delta$~=~44\arcdeg57\arcmin; epoch 1994).

We chose the HR5127 and Phi Herculis regions for low dust contrast
as determined by the IRAS 100~\micron\ map (\cite{whe94})
and the absence of known extragalactic radio sources (\cite{her92}).
The HR5127 observation lasted from 5.99~UT to 6.97~UT,
and the Phi Herculis observation lasted from 8.29~UT to 8.91~UT.
We calibrated the instrument before and after each observation
using a membrane transfer standard (\cite{fis92}).

We observed Jupiter from 4.86~UT to 4.95~UT to measure the
beam size.  We confirm the membrane calibration
using the best-fit beam size and the temperature of Jupiter
(\cite{gri86}).  We assume a 10\% uncertainty in calibration.
The calibration is such that a chopped beam centered
between sky regions with temperatures $T_1$ and $T_2$ would yield
$\Delta~T = T_1 - T_2$ in the absence of instrumental noise.

Anisotropy experiments are potentially susceptible to off-axis
response to local sources,
particularly the Earth, the balloon, and the Moon.
We measured the unchopped off-axis response in the 3.5 cm$^{-1}$
band to be $\geq$ 70~dB below the on-axis response at angles from
15\arcdeg\ to 25\arcdeg\ in elevation above the boresight.
We have not made comparable measurements of the chopped sidelobe
response in azimuth.
Our strongest argument against possible sidelobe contamination
is that the observed structure is stationary on the sky while
the horizon and local foregrounds shift.
Because the elevation of the observation changes as we track the target
star, we expect any sidelobe response from the Earth and the balloon
to change over time.
To minimize sidelobe response from the Moon,
we observe when the Moon is not full.
HR5127 was $\sim$ 56\arcdeg\ away from the Moon,
and Phi Herculis was $\sim$ 66\arcdeg\ away from the Moon.
The Moon set during the Phi Herculis observation.

\section{Data Analysis}
\subsection{Data Reduction}

We remove transients due to cosmic rays using an algorithm
described
by Alsop et al.(1992).  This procedure excludes approximately 15\%
of the data.
We demodulate the detector output using the sinusoidal reference
from the chopping secondary to produce antenna temperature differences
$\Delta T_A$ on the sky.  The demodulation produces a data set
in phase with the optical signal and a data set 90\arcdeg\
 out of phase
with the optical signal.  The noise averaged over the observations
gives respective CMB sensitivities of
590, 270, 620, and 4800 $\mu K \sqrt{s}$ for HR5127
and 600, 300, 630, and 4800 $\mu K \sqrt{s}$ for Phi Herculis
in the 3.5, 6, 9, and 14 cm$^{-1}$ bands.

The signal in each band is significantly offset from zero for both
HR5127 and Phi Herculis.
We attribute the offset to chopped emissivity differences on the
primary mirror and chopped atmospheric emission.
The averages of the measured instrumental offsets in antenna
temperature were 0.7, 0.2, 1.5, and 3.0~mK in the
3.5, 6, 9, and 14 cm$^{-1}$ bands.
The offset fluctuates in the higher frequency bands with
amplitudes of 200~$\mu$K and 300~$\mu$K at
9 and 14 cm$^{-1}$.
The period of the fluctuations is $\sim$ 300~s.
We subtract the offset and offset drift with a linear least
squares fit to each half scan from 0\arcdeg\ to +4\arcdeg\ to 0\arcdeg\
and from 0\arcdeg\ to -4\arcdeg\ to 0\arcdeg.
Each half scan takes 72~s.
This method removes the $\sim$ 300~s fluctuations as long as the $\sim$ 300~s
fluctuations are not phase-synchronous with the scan period.
After the offset
and offset drift removal, the residual temporal correlation
between frequency bands is a few percent.

For each observation we calculate the means and 1 $\sigma$
uncertainties of the antenna temperature differences for 29 pixels
separated by 17\arcmin\ on the sky.
Figure~\ref{hrant} shows the measured antenna
temperature differences for the HR5127 observation.  There is
significant signal which is not consistent with Gaussian random
instrument noise.  The reduced $\chi^2$ for the null hypothesis are
64/29, 94/29, 49/29, and 31/29 in the
3.5, 6, 9, and 14~cm$^{-1}$ bands.
By comparison, the out of phase components are mostly consistent
with Gaussian random instrument noise.  (Reduced $\chi^2$ = 20/29,
18/29, 36/29, and 27/29 at 3.5, 6, 9, and 14~cm$^{-1}$)  There is
correlated structure in the 3.5, 6, and 9~cm$^{-1}$ bands and no
significant signal at 14~cm$^{-1}$ as expected for CMB anisotropy.  If
we divide the observation into two equal time periods, the
differences between the two halves are consistent with zero in the
main CMB bands.
(Reduced $\chi^2$ = 20/29, 26/29, and 28/29 at 3.5,
6, and 9~cm$^{-1}$)  The temporal stability of the signal in each pixel
suggests that the signal is not due to either sidelobe response or
atmospheric emission.

Figure~\ref{phant} shows the measured antenna
temperature differences for the Phi Herculis observation.  There is
significant signal which is not consistent with Gaussian random
instrument noise.  The reduced $\chi^2$ for the null hypothesis are
49/29, 156/29, 86/29, and 77/29 in the
3.5, 6, 9, and 14~cm$^{-1}$ bands.
By comparison, the out of phase components are mostly consistent
with Gaussian random instrument noise.
(Reduced $\chi^2$ = 242/29,
38/29, 15/29, and 35/29 at 3.5, 6, 9, and 14~cm$^{-1}$)
In addition to the correlated structure at 3.5, 6, and 9~cm$^{-1}$
there is structure at 14~cm$^{-1}$.  If
we divide the observation into two equal time periods, the
differences between the two halves are consistent with zero in the
3.5 and 6~cm$^{-1}$ bands. (Reduced $\chi^2$ = 21/29 and 22/29,
respectively)  The temporal stability of the signal in each pixel
at 3.5 and 6~cm$^{-1}$
suggests that the signal is not due to either sidelobe response or
atmospheric emission.  We interpret the structure at 14~cm$^{-1}$ as
atmosphere because of its time variability.

\subsection{Spectral Discrimination}
In order to test the hypothesis than the signals in all frequency
bands originate from a single morphology,
we determined a best-fit sky model $y_i$ by minimizing
\begin{equation}
\chi^2 = \sum^{4}_{j=1}\sum^{29}_{i=1} (x_{ij} - a_jy_i)^2/
	\sigma^{2}_{ij},\label{chi2}
\end{equation}
where $x_{ij}$ and $\sigma_{ij}$ are the measured mean and
uncertainty in angular bin i for frequency band j.
The $a_j$ denote scale factors for different astrophysical processes.
Table~\ref{tabspec} gives reduced $\chi^2$ using scale factors $a_j$
for CMB, free-free, synchrotron, and dust emission.
The 10\% calibration uncertainty is included in
calculating the reduced $\chi^2$.
The HR5127 data set is spectrally consistent with free-free,
synchrotron, and CMB
emission.
The complete Phi Herculis data set is spectrally inconsistent with
all four
astrophysical processes.
However, the Phi Herculis data at 3.5~cm$^{-1}$ and
6~cm$^{-1}$
are spectrally inconsistent with synchrotron emission and spectrally
consistent with free-free, dust, and CMB emission

\subsection{Galactic and Extragalactic Emission}
At 6, 9, and 14~cm$^{-1}$,
the predominant astrophysical source of confusion
is thermal emission from interstellar dust.
Using a brightness scaling for high-latitude dust
(\cite{mei93a,fis95}), we can extrapolate the IRAS 100~\micron\ data
(\cite{whe94}) to our frequency bands.
At most, we expect the differential dust emission
to contribute a few~$\mu$K in antenna
temperature at 14~cm$^{-1}$ and negligibly at lower frequencies
in both HR5127 and Phi Herculis.
The observed 14~cm$^{-1}$ structure in Phi Herculis does not
correlate with the structure obtained by convolving our
beam and scan pattern with the IRAS 100~\micron\ map.

A catalog search (\cite{her92}) shows no bright radio sources
in either region.  We can also set conservative amplitude limits
on free-free and synchrotron emission with the
408~MHz Haslam et al. (1982) map.
If we convolve our scan pattern with the 30\arcmin $\times$
30\arcmin\ smoothed version of the map and assume antenna
temperature scales as $\Delta T_{A} \propto \nu^{\beta}$,
we obtain $\lesssim$ 1\% of the observed structure at
3.5~cm$^{-1}$ for synchrotron emission ($\beta$~=~-2.7)
and $\lesssim$ 10\% of the observed structure at
3.5~cm$^{-1}$ for free-free emission ($\beta$~=~-2.1).

\subsection{Atmosphere}
During the Phi Herculis observation the balloon altitude
dropped to 32~km and fluctuated
with an amplitude of several hundred feet every $\sim$ 300~s.
If we process the balloon altitude like a detector signal, removing
offset and offset drift for each half scan and calculating a mean for
each azimuth bin, we find that the 14~cm$^{-1}$ structure in
Phi Herculis correlates (-0.68, null probability = 5$\times10^{-5}$)
with the binned balloon
altitude.
This strong correlation suggests that the observed structure
is due to atmospheric contamination.
We expect to observe less
atmospheric emission during the HR5127 observation
because the average balloon altitude was 35~km.

If all the structure at 14~cm$^{-1}$ is attributed to modeled
atmospheric
contamination, atmospheric emission contributes less than
5\% of the observed structure at 3.5~cm~$^{-1}$ and 10\% of the observed
structure at 6~cm$^{-1}$.
Additionally, the structure in the lower
frequency bands is less correlated
with the binned balloon altitude
(-0.09, null probability~=~0.64 at 3.5~cm$^{-1}$;
-0.17, null probability = 0.39 at 6 cm$^{-1}$).
In our CMB analysis of Phi Herculis,
we consider only the 3.5~cm$^{-1}$ and
6~cm$^{-1}$ data.
An alternative analysis which considers the linear
combinations of all frequency
band data that are orthogonal to the atmospheric contamination yields
the same CMB anisotropy
confidence intervals within 5\%.
Similar analyses have been used to separate CMB from Galactic
and extragalactic foreground sources (\cite{dod94}).

\section{Discussion}
We model the anisotropy $\Delta T(x)$ with a correlation function
\begin{equation}
\langle\Delta\rm T(\bf{x_1})\rm\Delta T(\bf{x_2})\rm\rangle=C_0 c(\theta),
\end{equation}
using a likelihood ratio statistic to set limits on the amplitude of
the fluctuations $\sqrt{C_0}$ (\cite{che94}).
We include
correlations between frequency bands in the covariance matrix.
However, the off diagonal elements affect the calculated values
of $\sqrt{C_0}$ by $\lesssim$ 2\%.

To set limits on the CMB anisotropy for the HR5127 observation,
we include the 3.5, 6, and 9 cm$^{-1}$ data.  For the Phi Herculis
observation, we include only the 3.5 and 6~cm$^{-1}$ data.
Assuming a Gaussian 0\fdg5 FWHM beam,
we present two separate measures of the fluctuations in
Table~\ref{tabcmb} for each observation.
We find the most probable $\sqrt{C_0}$ for a Gaussian correlated sky
$c(\theta) = \exp(-\theta^2/2\theta^2_c)$ where
$\theta_c$ = 25\arcmin\ (GACF).
We also find a flat band power estimate
$\langle l(l+1)C_l/2\pi\rangle^{1/2}$
for
$c(\theta) = \frac{1}{4\pi}\sum~(2l+1)~C_l~P_l(\cos\theta)$
where
$C_l \propto 1/l(l+1)$.

For comparison, Table~\ref{taboldcmb} lists the most probable
values of CMB anisotropy from the fourth flight of MAX.
The GACF limits are as much as 20\% lower than previously published
(\cite{cla94,dev94}).
The previous MAX4 analyses did not include sufficient power in the
off diagonal elements and improperly normalized the larger beam
sizes in the theoretical covariance matrix.
The distributed data sets, however, have not changed.
These revised MAX4
lower values are closer to the values measured in
the two scans reported here and for the region
of sky near the star Mu Pegasi (\cite{mei93a}).

Bayesian analyses and likelihood ratio analyses
of the MAX4 and MAX5 data sets yield similar most
probable values of $\Delta T/T$ with the exception of the Sigma
Herculis data set.
A Bayesian analysis of the Sigma Herculis data yields the 68\%
confidence interval GACF
$\Delta T/T = 4.0^{+1.7}_{-1.1}\times 10^{-5}$.
A likelihood ratio analysis yields the 68\% confidence interval
GACF
$\Delta T/T = 2.8^{+1.2}_{-0.9}\times 10^{-5}$
as shown in Table~\ref{taboldcmb}.
The discrepancy between the two analyses
is being investigated.

\section{Conclusion}
We have presented new MAX measurements of CMB anisotropy
with high sensitivity at 0\fdg5 angular scales.
The amplitude and morphology of the observed structure is not
consistent with emission from known forms of interstellar dust,
synchrotron,
and free-free emission.
The HR5127 data are consistent with CMB anisotropy
with flat band power
$\Delta T/T = 1.2^{+0.4}_{-0.3}\times 10^{-5}$.
The Phi Herculis data are consistent with CMB anisotropy if we
either consider the linear combinations of the data that are
independent of the spectrum of atmospheric emission
or only use the 3.5 and 6~cm$^{-1}$ data.
For a flat band power estimate
$\Delta T/T = 1.9^{+0.7}_{-0.4}\times 10^{-5}$ for Phi Herculis.
The upper and lower
limits represent a 68\%\ confidence interval
added in quadrature with a 10\%\
calibration uncertainty.
The difference between the HR5127 and Phi Herculis results is similar to
the difference between the previously reported results for the sky
regions near Mu Pegasi (\cite{mei93a})
and Gamma Ursae Minoris (\cite{als92,gun93,dev94}).
The HR5127 and Phi Herculis data are available from the authors.
\acknowledgments

This work was supported by the National Science Foundation
through the Center for Particle Astrophysics
(cooperative agreement AST~91-20005),
the National Aeronautics and Space Administration under grants
NAGW-1062 and FD-NAGW-1221,
the University of California,
and previously the California Space Institute.
N. Figueiredo was partially supported by
Conselho Nacional de Desenvolvimento
Cient\'{\i}fico e Tecnol\'{o}gico, Brazil.
We would like to thank O.~Levy for his assistance with
the flight preparation, the MSAM team for borrowed
equipment and discussions, and M. White.

\clearpage
\begin{deluxetable}{crrrr}
\footnotesize
\tablecaption{
Spectral constraints:  reduced $\chi^2$ from Equation 1,
using $a_j$ for different astrophysical
processes. \label{tabspec}}
\tablewidth{0pt}
\tablehead{
\colhead{Observation}
& \colhead{CMB}
& \colhead{Free-free\tablenotemark{a}}
& \colhead{Synchrotron\tablenotemark{b}}
& \colhead{Dust\tablenotemark{c}}
}
\startdata
HR5127 &97/87 &92/87 &100/87 &136/87\nl
Phi Herculis &119/87 &132/87 &164/87 &170/87\nl
Phi Herculis (3.5 and 6~cm$^{-1}$ only) &25/29 &36/29&50/29 &29/29\nl
\enddata

\tablenotetext{a}{Assumes brightness $I_{\nu} \propto \nu^{-0.1}$}
\tablenotetext{b}{Assumes $I_{\nu} \propto \nu^{-0.7}$}
\tablenotetext{c}{Assumes $I_{\nu} \propto \nu^{1.4} B_{\nu}(T=18K)$
	where $B_{\nu}$ is the Planck function}

\end{deluxetable}

\clearpage
\begin{deluxetable}{crr}
\footnotesize
\tablecaption{MAX5 limits on CMB anisotropy.
	(Upper and lower limits represent 68\% confidence interval
	added in quadrature with 10\% calibration uncertainty.)
	\label{tabcmb}}
\tablewidth{0pt}
\tablehead{
\colhead{Observation}
& \colhead{GACF $\Delta T/T_{CMB}$}
& \colhead{Flat Band $\langle\frac{l(l+1)C_l}{2\pi}\rangle^{1/2}$}
}
\startdata
HR5127
		&$1.8_{-0.5}^{+0.6} \times 10^{-5}$
		&$1.2_{-0.3}^{+0.4} \times 10^{-5}$\nl
Phi Herculis
		&$2.9_{-0.7}^{+1.0} \times 10^{-5}$
		&$1.9_{-0.4}^{+0.7} \times 10^{-5}$\nl
\enddata
\end{deluxetable}

\clearpage
\begin{deluxetable}{crr}
\footnotesize
\tablecaption{Reanalyzed MAX4 limits on CMB anisotropy.
(Upper and lower limits represent 68\% confidence interval
added in quadrature with 10\% calibration uncertainty.)\label{taboldcmb}}
\tablewidth{0pt}
\tablehead{
\colhead{Observation}
& \colhead{GACF $\Delta T/T_{CMB}$}
& \colhead{Flat Band $\langle\frac{l(l+1)C_l}{2\pi}\rangle^{1/2}$}
}
\startdata
GUM 		&$3.1_{-0.8}^{+1.2} \times 10^{-5}$
			&$2.0_{-0.4}^{+0.6} \times 10^{-5}$  \nl
Iota Draconis 	&$2.6_{-0.8}^{+1.3} \times 10^{-5}$
			&$1.7_{-0.5}^{+0.8} \times 10^{-5}$  \nl
Sigma Herculis
			&$2.8_{-0.9}^{+1.2} \times 10^{-5}$
			&$1.8_{-0.6}^{+0.8} \times 10^{-5}$ \nl
\enddata
\end{deluxetable}

\clearpage

\figcaption{
Antenna temperature differences ($\pm 1 \sigma$) for the data
near HR5127.  The points are separated by 17\arcmin\ in azimuth.
The solid lines show the response to a point source in each
of the bands.
\label{hrant} }

\figcaption{
Antenna temperature differences ($\pm 1 \sigma$) for the data
near Phi Herculis.
The points are separated by 17\arcmin\ in azimuth.
The solid lines show the response to a point source in each
of the bands.
\label{phant} }

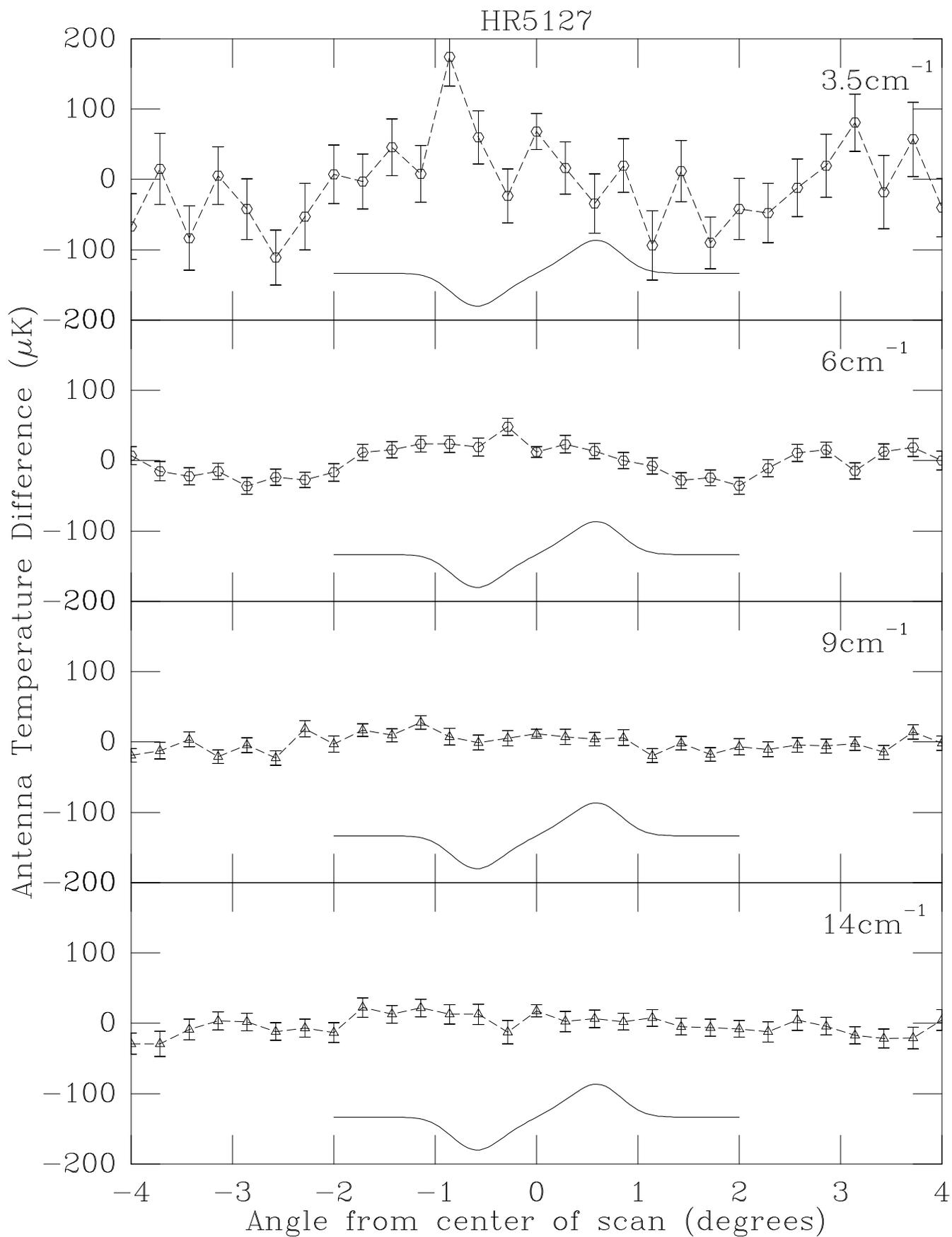
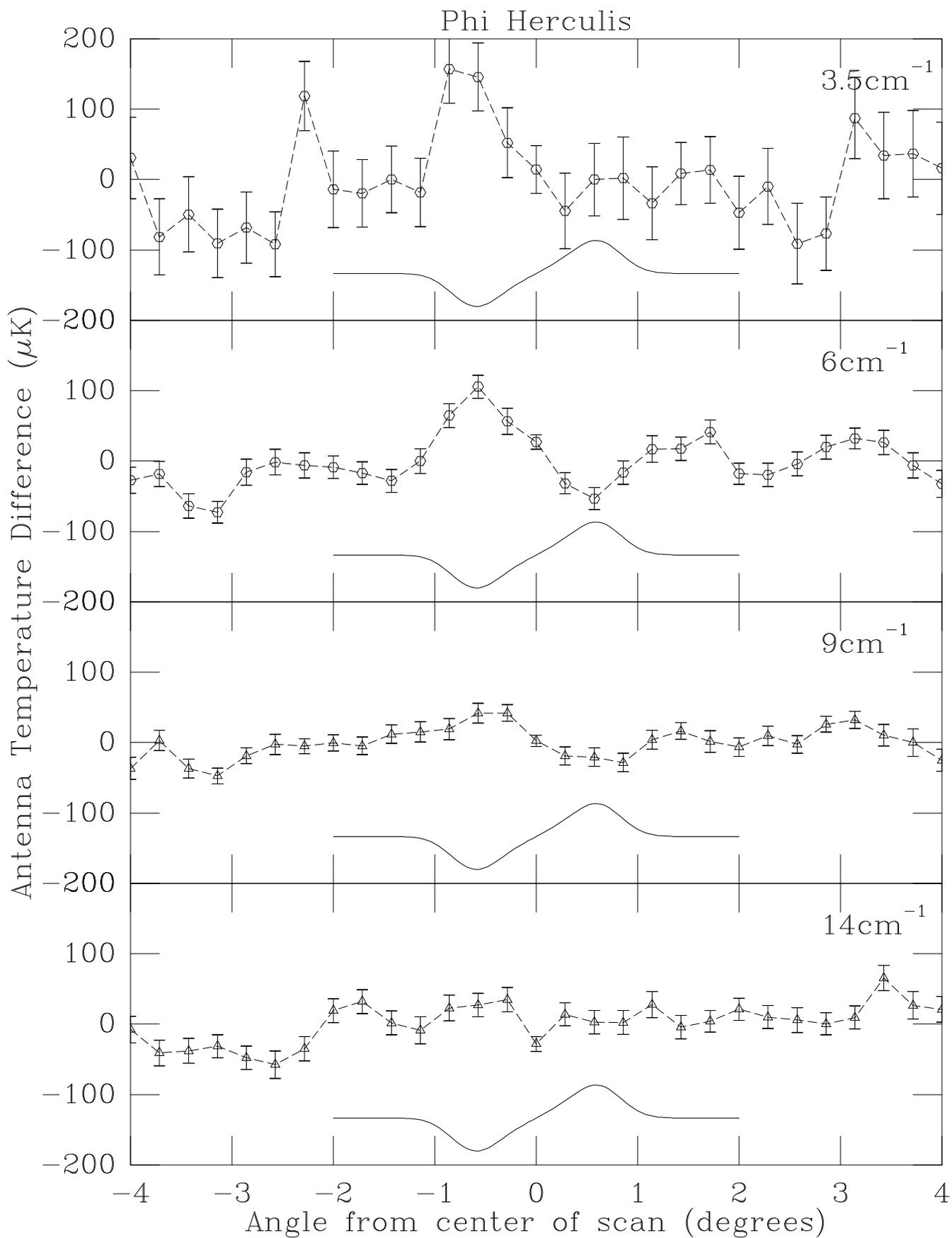
\end{document}